\documentclass{jpsj3}
\usepackage{txfonts,braket}
\usepackage[dvipdfmx]{xcolor}

\title{Piezomagnetic effect of a rare-earth-based altermagnet $\rm TbPt_{6}Al_{3}$}

\author{Ryohei Oishi$^{1,2}$\thanks{oishi@es.hokudai.ac.jp}, Kazunori Umeo$^3$, Takuya Aoyama$^{2,4}$,Takahiro Onimaru$^2$, and Kaya Kobayashi$^1$}
\inst{$^1$Research Institute for Electronic Science, Hokkaido University, Sapporo 001-0020, Japan\\
$^2$Department of Quantum Matter, Graduate School of Advanced Science and Engineering, Hiroshima University, Higashihiroshima, Hiroshima 739-8530, Japan\\
$^3$Low Temperature Experiment Section, Integrated Experimental Support and Research Department, Division for Technical Development and Support Management, Core Facility Management Center, Hiroshima University, Higashihiroshima, Hiroshima 739-8526, Japan\\
$^4$International Institute for Sustainability with Knotted Chiral Meta Matter (WPI-SKCM$^2$), Hiroshima University, Higashi-Hiroshima, Hiroshima 739-8531, Japan}

\abst{We have investigated the piezomagnetic (PZM) effect of the rare-earth-based $g$-wave altermagnet $\rm TbPt_{6}Al_{3}$ by magnetization measurements of single-crystalline samples under uniaxial stress $\sigma$.
The magnetization in magnetic field along the trigonal $a$ axis increases linearly with $\sigma$ for $T$ $\le$ $T_{\rm N}$, indicating the emergence of PZM effect, while the theoretically predicted nonlinear PZM effect was not observed.
PZM coefficient of $Q_{11}$ at 2 K is obtained as $9.1 \times 10^{-3} \rm \mu_{B}/(f.u.\,MPa)$, which is larger by more than two orders of magnitude than those for other altermagnets and noncollinear antiferromagnets.
Temperature dependence of $Q_{11}$ below $T_{\rm N}$ yielded the critical component $\beta$ as 0.28, whose value is close to that of the magnetic moment estimated by the neutron powder diffraction.
We propose that the large $Q_{11}$ and the large poling field of 10000 Oe to achieve the single-domain state in $\rm TbPt_{6}Al_{3}$ are due to the strong relativistic spin-orbit coupling of the 4$f$ electrons in the Tb$^{3+}$ ions.}

\begin{document}
\maketitle
Time-reversal symmetry (TRS) breaking in a magnetic point group induces functional phenomena such as the anomalous Hall effect, spin-splitting, and the piezomagnetic (PZM) effect\cite{Spintronics1,Spintronics2,Spintronics3,Spintronics4,Naka,JPSJ88.123702}.
These phenomena have been investigated for ferromagnets and non-collinear antiferromagnets (AFMs).
Recently, a classification using the spin Laue group gives a spotlight on the collinear AFMs with TRS breaking, which is referred to as an altermagnet (AM)\cite{Hariki,Smejkal}.
Even in the absence of the relativistic spin-orbit coupling (SOC), AMs exhibit spin-split bands.
Depending on the symmetry of the momentum-dependent spin splitting, AMs are characterized as $d$-, $g$-, and $i$-wave types\cite{AM_PRX1,AM_PRX2}.
From the perspective of multipoles, the symmetry classification indicates that AMs are characterized by orderings of the magnetic octupole and magnetic toroidal quadrupole\cite{Multipole}.

Among TRS breaking phenomena, the linear PZM effect is defined by a linear coupling between the magnetization $M_{i}$ and the external stress tensor $\sigma_{jk}$, as $M_{i} = Q_{ijk}\sigma_{jk}$.
Here, a third-rank axial-$c$ tensor $Q_{ijk}$ is a PZM tensor, and the corresponding free energy is given by $F = Q_{ijk}\sigma_{jk}H_{k}$\cite{PME_theory1,PME_theory2}.
The microscopic mechanism of the linear PZM effect was theoretically investigated by including the relativistic SOC\cite{CoF2_1,CoF2_2}.
For AMs, on the other hand, the unconventional PZM effect has been discussed in the absence of the relativistic SOC\cite{JPSJ_Ogawa,JPSJ_Naka}.
For example, $g$-wave AMs in a two-dimensional tetragonal system exhibit the nonlinear PZM effect, in which the induced magnetization is quadratic in $\sigma$.
To confirm those predictions and to achieve a deeper understanding of AMs, experimental investigations of the PZM effect on other AMs are needed.

MnTe with a hexagonal NiAs-type crystal structure has been studied theoretically and experimentally, and is now established as a representative AM\cite{MnTe,AM_PRX1,AM_PRX2,INS_MnTe}.
The nonrelativistic spin splitting was observed by angle-resolved photoemission spectroscopy experiments, which determined that MnTe is the $g$-wave type AM\cite{ARPES_MnTe}.
Furthermore, the linear PZM effect was reported by magnetization measurements on polycrystalline samples under uniaxial stress\cite{PZM_MnTe}.
The magnetization below $T_{\rm N}$ increases linearly with $\sigma$, yielding an averaged $Q$ of $1.38 \times 10^{-8} \mu_{\rm B}/(\rm f.u.\,MPa)$ at 300 K.

\begin{figure*}[t]
	\begin{center}
	\includegraphics[width=\linewidth]{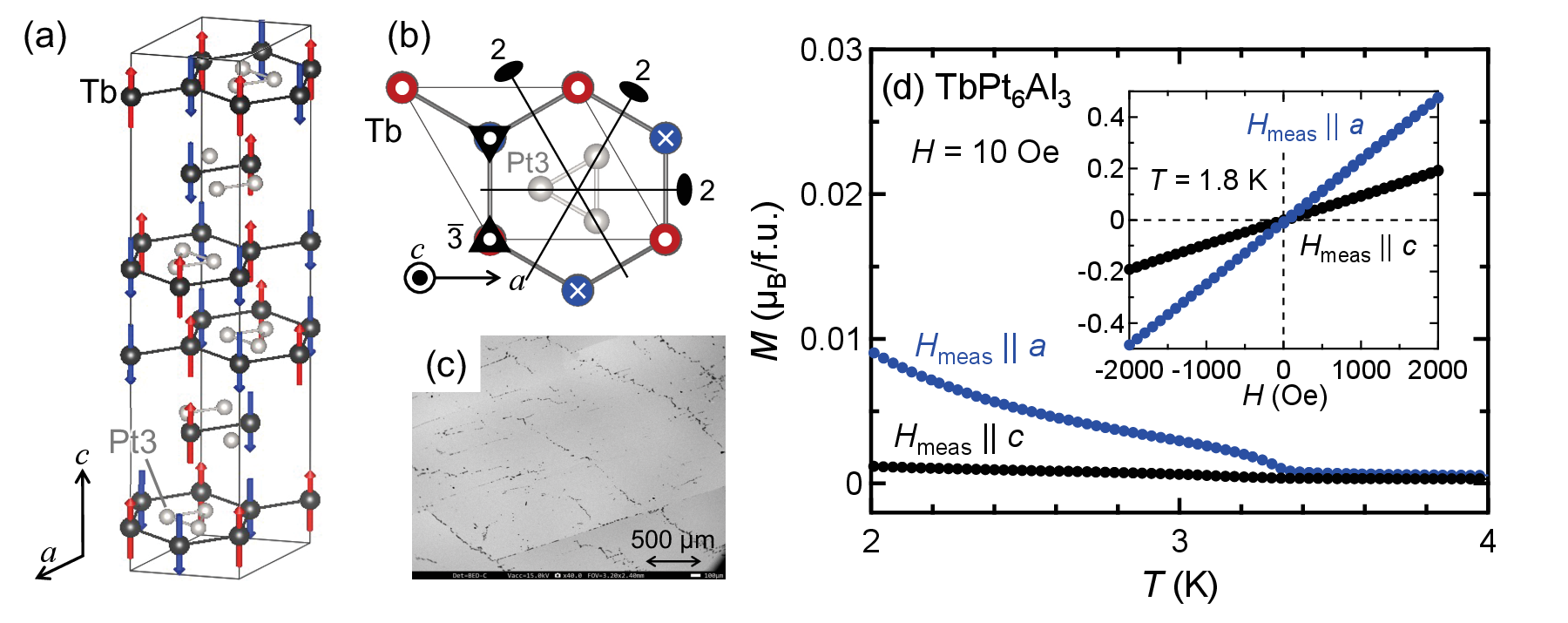}
	\caption{(Color online) (a) Magnetic structure of $\rm TbPt_{6}Al_{3}$~\cite{TbPt6Al3} in which Pt1, Pt2, Pt4, and Al are omitted for clarity. (b) The $\rm Tb_{2}Pt_{3}$ honeycomb layer viewed from the $c$ axis. The Tb atoms have $\bar{3}$ symmetry. The opposite magnetic moments on the honeycomb plane are connected by the two-fold rotational symmetry. (c) Backscattered electron image of $\rm TbPt_{6}Al_{3}$. The impurity phase shown by black dots is PtAl. (d) The temperature dependence of magnetization $M(T)$ at 10 Oe for $H_{\rm meas} \parallel a$ and $H_{\rm meas} \parallel c$. The inset shows the isothermal magnetization $M(H)$ at 1.8 K for $H_{\rm meas} \parallel a$ and $H_{\rm meas} \parallel c$.}
	\label{f1}
	\end{center}
\end{figure*}

A series of $R\rm Pt_{6}Al_{3}$ ($R$ = Ce, Pr, Nd, Sm, Gd, and Tb), which crystallizes in the $\rm NdPt_{6}Al_{3}$-type trigonal structure with the space group of $R\bar{3}c$\cite{RPt6Al3}, is one of the candidates of the rare-earth-based AMs.
As shown in Fig. \ref{f1}(b), the $R$ site has a $\bar{3}$ symmetry and the nearest neighbor $R$ atoms are connected by a two-fold rotational symmetry.
The nonsymmorphic symmetry between the $R$ sites is the key ingredient to breaking TRS in $R\rm Pt_{6}Al_{3}$.
The physical property measurements on single crystals revealed that $R\rm Pt_{6}Al_{3}$ ($R$ = Nd, Sm, Gd, and Tb) have antiferromagnetic orders\cite{NdPt6Al3,SmPt6Al3,GdPt6Al3,TbPt6Al3}.

Among the $R\rm Pt_{6}Al_{3}$ series, $\rm TbPt_{6}Al_{3}$ was reported as a rare-earth-based $g$-wave AM by neutron powder diffraction experiments\cite{TbPt6Al3}.
Figure \ref{f1}(a) shows a collinear AFM structure of $\rm TbPt_{6}Al_{3}$, whose magnetic point group was determined as $\bar{3}m.1$.
Due to the strong relativistic SOC of the 4$f$ electrons, the ordered moments of the $\rm Tb^{3+}$ ion point along the $c$ axis.
As shown in Fig. \ref{f1}(b), the opposite sublattices are antiferromagnetically coupled in the honeycomb plane.
In this magnetic point group with broken TRS, the PZM tensor is given as
\begin{equation}
Q_{k\mu} = 
	\begin{pmatrix}
	Q_{11} & -Q_{11} & 0 & Q_{14} & 0 & 0 \\
	0 & 0 & 0 & 0 & -Q_{14} & -2Q_{11} \\
	0 & 0 & 0 & 0 & 0 & 0
	\end{pmatrix}
\label{eq1}
\end{equation}
, where $k$ is the direction of applied magnetic fields ($k$ = 1 and 3 corresponding to the $a$ and $c$ axes)\cite{bilbao}.
Using Voigt notation, the stress tensor $\sigma_{jk}$ is written using a single suffix form $\sigma_{\mu}$ ($\mu$ = 1, 2, 3, 4, 5, and 6).
Therefore, $\sigma_{1}$ and $\sigma_{3}$ represent the uniaxial stresses along the $a$ and $c$ axes, respectively.
The finite value of $Q_{11}$ indicates that the uniaxial stress applied along the $a$ axis induces the magnetization in the $a$ axis direction.

In this Letter, we reported the emergence of the linear PZM effect in $\rm TbPt_{6}Al_{3}$ by magnetization measurements with single-crystalline samples under the uniaxial stress along the $a$ axis.
As a result, $Q_{11}$ at 2 K was estimated as $9.1 \times 10^{-3} \rm \mu_{B}/(f.u.\,MPa)$, whose value is larger than that of the typical $g$-wave AM MnTe at 300 K.
The nonlinear PZM effect predicted theoretically\cite{JPSJ_Ogawa}, on the other hand, was not observed in $g$-wave AM $\rm TbPt_{6}Al_{3}$.
Our results highlighted the strong relativistic SOC of the 4$f$ electrons as a key ingredient to produce the large linear PZM effect.

The single-crystalline sample of $\rm TbPt_{6}Al_{3}$ was grown by the Czochralski method\cite{TbPt6Al3}.
As shown in Fig. \ref{f1}(c), an impurity phase of PtAl was detected by the electron-probe microanalysis, causing the slight deficiency of Pt and Al in $\rm TbPt_{6}Al_{3}$ phase.
For the magnetization measurements, the crystals were oriented parallel to the trigonal $a$ and $c$ axes by the back-reflection Laue X-ray method.
The oriented crystals were cut into appropriate dimensions by spark erosion.

The uniaxial stresses $\sigma_{1}$ and $\sigma_{3}$ were applied to platelike samples with 0.5 mm thickness and a mass of 50 mg using a home-made pressure cell made of ZrO$_{2}$.
The applied pressure was estimated with the hydraulic press force and the cross-sectional area, in addition to the superconducting transition temperatures of tin under the hydrostatic pressure\cite{TC_tin}.
To keep the hydrostatic pressure of tin used as a monometer, the tin was covered by teflon sheet\cite{KUmeo_UP}.
For the temperature-dependent magnetization measurements $M(T)$, the pressure cell was inserted into a Quantum Design MPMS.
$M(T)$ measurements were carried out both with and without samples.
By subtracting the data without the sample from that with the sample\cite{SQUID_sub}, we obtained the signal of $\rm TbPt_{6}Al_{3}$.

$\rm TbPt_{6}Al_{3}$ has two inequivalent AFM domains associated with TRS breaking, and each domain has the opposite sign of $Q_{k\mu}$ tensor.
By cooling the sample below $T_{\rm N}$ = 3.5 K under $\sigma_{1}$ at a poling field $H_{\rm pol}$\cite{PZM_MnTe}, a single domain state is achieved.
After the poling process, the data of $M(T)$ were measured in a heating process from 2 K in the magnetic field $H_{\rm meas}$.

\begin{figure}
\includegraphics[width=\linewidth]{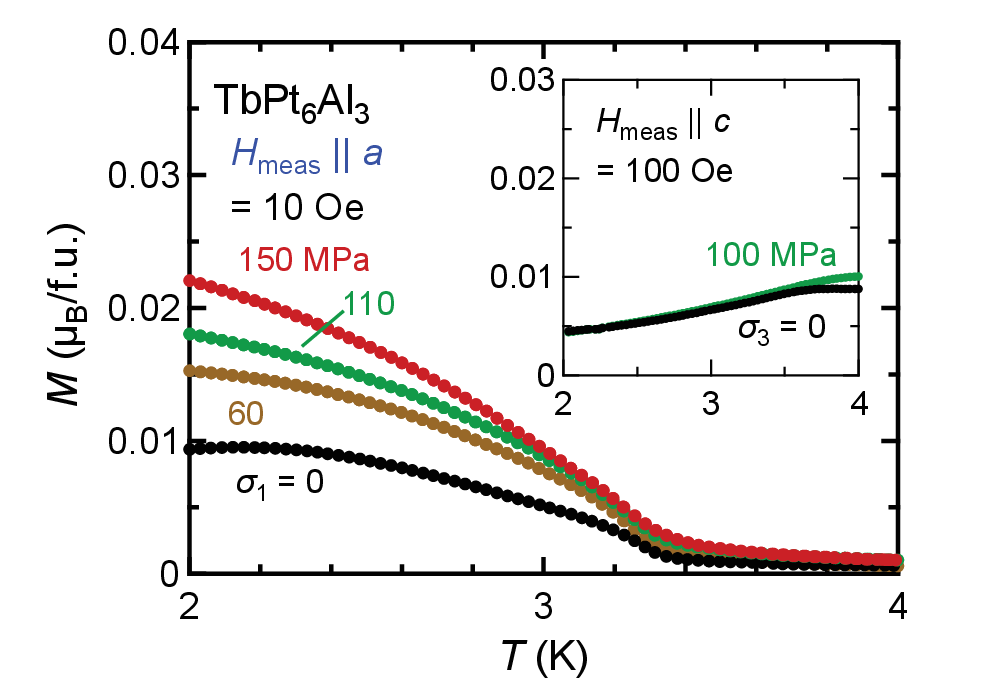}
\caption{(Color online) Temperature dependence of magnetization $M(T)$ of the single crystal of $\rm TbPt_{6}Al_{3}$ under $\sigma_{1}$. 
Measurements were carried out after the poling process at $H_{\rm pol}$ = 1000 Oe.
The inset shows the $M(T)$ data under $\sigma_{3}$.}
\label{f2}
\end{figure}

The temperature dependences of $M$ in $\rm TbPt_{6}Al_{3}$ at a low magnetic field of 10 Oe under $H_{\rm meas} \parallel a$ and $H_{\rm meas} \parallel c$ are displayed in Fig. \ref{f1}(d), whose data were measured using a translucent plastic straw after the poling process.
In the previous study, $M(T)/H$ for $H_{\rm meas} \parallel a$ = 1000 Oe was reported to become flat below $T_{\rm N}$, which is consistent with the Tb moment pointing along the $c$ axis as shown in Fig. \ref{f1}(a)\cite{TbPt6Al3}.
The $M(T)$ data at $H_{\rm meas} \parallel a$ = 10 Oe, on the other hand, shows a small spontaneous moment of 0.009 $\mu_{\rm B}$/f.u., whose value is much smaller than 5 $\mu_{\rm B}$/f.u. for the ordered moment determined by neutron powder diffraction\cite{TbPt6Al3}.
For $H_{\rm meas} \parallel c$ = 10 Oe, $M(T)/H$ becomes almost flat below $T_{\rm N}$.
As shown in Fig. \ref{f1}(d) inset, no hysteresis is observed in the isothermal magnetization for the $a$ and $c$ axes.
Note that the magnetic component in the $a$ axis is zero under the magnetic point group of $\bar{3}m.1$ for $\rm TbPt_{6}Al_{3}$.
Thus, the spontaneous moment on the honeycomb plane may be attributed to a slight canting of the ordered moment from the $c$ axis or an intrinsic PZM effect due to the local disorder of Pt and Al.
The magnitude of the spontaneous moment for the multiple samples remains within 15\% of 0.009 $\mu_{\rm B}$/f.u.

\begin{figure}
\includegraphics[width=\linewidth]{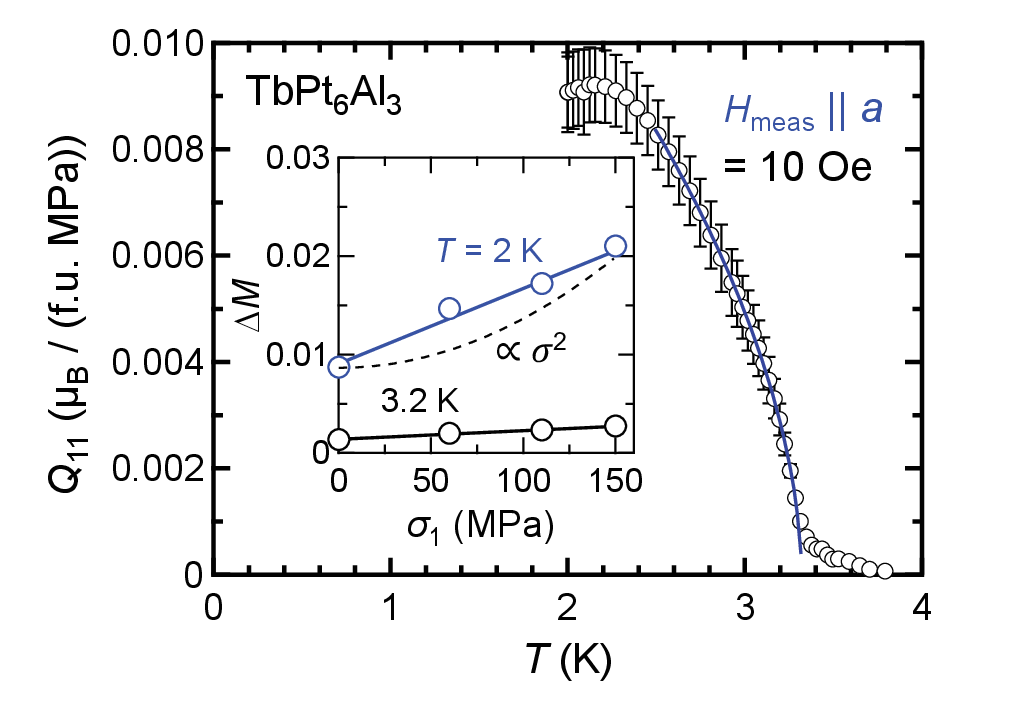}
\caption{(Color online) Temperature dependence of the piezomagnetic coefficient $Q_{11}$ of $\rm TbPt_{6}Al_{3}$. The blue solid line represents a fit with an equation of $Q_{0}[(T_{\rm N}-T)/T_{\rm N}]^{2\beta}$ (see text). The inset shows the stress dependence of $\Delta M = M(T) - M(4\,\rm K)$ at $T$ = 2 K and 3.2 K. The solid lines are fits with the linear function. The nonlinear PZM effect predicted in ref \citen{JPSJ_Ogawa} is represented by a dashed line.}
\label{f3}
\end{figure}

Figure \ref{f2} shows the temperature dependence of $M(T)$ at $H _{\rm meas}$ = 10 Oe under the uniaxial stress applied along the $a$ axis.
At $\sigma_{1}$ = 0, $M(T)$ shows a spontaneous moment of 0.01 $\mu_{\rm B}$/f.u. below $T_{\rm N}$, which is also observed in Fig. \ref{f1}(d).
With increasing $\sigma_{1}$ to 150 MPa, the spontaneous moment at 2 K is monotonically enhanced from 0.009 $\mu_{\rm B}$/f.u. to 0.022 $\mu_{\rm B}$/f.u.
In the whole range of $\sigma_{1}$, $T_{\rm N}$ remains unchanged.

We subtracted the value of $M$ at 4 K from that below 4 K to estimate the relative change from the paramagnetic state.
The subtracted data $\Delta M = M(T) - M(4\,\rm K)$ at 2 K and 3.2 K are plotted against $\sigma_{1}$ in the inset of Fig. \ref{f3}.
Note that $\Delta M$ increases linearly with $\sigma_{1}$, which is the experimental evidence of the linear PZM effect.
Fitting the data of $\Delta M$ with the linear function yields $Q_{11}$ at 2 K as $9.1 \times 10^{-3} \mu_{\rm B}/(\rm f.u.\,MPa)$.
As shown in Fig. \ref{f2}, magnetization below $T_{\rm N}$ for the $c$ axis does not change when $\sigma_{3}$ is increased from 0 to 100 MPa, which is consistent with $Q_{33}$ = 0 in the PZM tensor for $\rm TbPt_{6}Al_{3}$ (Eq. (\ref{eq1})).

By calculating $Q_{11}$ value at various temperatures, we obtained the temperature dependence of $Q_{11}$ shown in Fig. \ref{f3}.
The blue solid line represents a fit to the data below $T_{\rm N}$ using an equation of $Q_{0}[(T_{\rm N}-T)/T_{\rm N}]^{2\beta}$, which gives $T_{\rm N}$ = 3.32(1) K and the critical component $\beta$ = 0.28(1).
The obtained critical component of $Q_{11}$ is close to 0.29(6) for the refined magnetic moment\cite{TbPt6Al3}.
The agreement of both $\beta$ values is consistent with the Landau theory for AMs\cite{Landau_AM1, Landau_AM2}.

Let us discuss the origin of the large magnitude of $Q_{11}$ = $9.1 \times 10^{-3} \rm \mu_{B}/(f.u.\,MPa)$ at 2 K for $\rm TbPt_{6}Al_{3}$, which is 100 times larger than those for transition-metal-based materials obtained by the magnetization measurements under pressure\cite{PZM_MnTe,PZM_MnF2_CoF2,PZM_Y2Ir2O7}.
The microscopic mechanisms have been proposed for the linear PZM effect, including $g$-tensor anisotropy and single-ion anisotropy\cite{CoF2_1}.
Each of them is proportional to either the first or the second order of the relativistic SOC constant.
Based on this mechanism, the large $Q$ value for $\rm TbPt_{6}Al_{3}$ is understood by the strong SOC of the 4$f$ electrons, resulting in Ising-type and localized magnetic moments.
Similarly, uranium-based ferromagnet URhGe shows the large PZM effect due to the strong SOC of the 5$f$ electrons\cite{PZM_URhGe}.
Moreover, the local inversion symmetry breaking at the rare-earth site and the midpoint between nearest-neighbor rare-earth ones in the honeycomb network gives rise to asymmetric SOC such as Dzyaloshinskii-Moriya (DM) interaction\cite{Moriya}.
Since the DM vector of $R\rm Pt_{6}Al_{3}$ is pointing along the $c$ axis\cite{SmPt6Al3}, the magnetic moments along the $a$ axis are canted, resulting in a ferromagnetic component in the honeycomb plane.
This scenario may also be applicable to the magnetic components of the $a$ axis of $\rm TbPt_{6}Al_{3}$ induced by the linear PZM effect.
To further understand the role of relativistic SOC in $Q$, inverse PZM effect of $\rm TbPt_{6}Al_{3}$ and tuning the SOC constant by chemical substitutions are essential\cite{Ce3Bi4(PtPd)3,Ce(PtPd)6Al3}.

\begin{figure}
	\begin{center}
	\includegraphics[width=\linewidth]{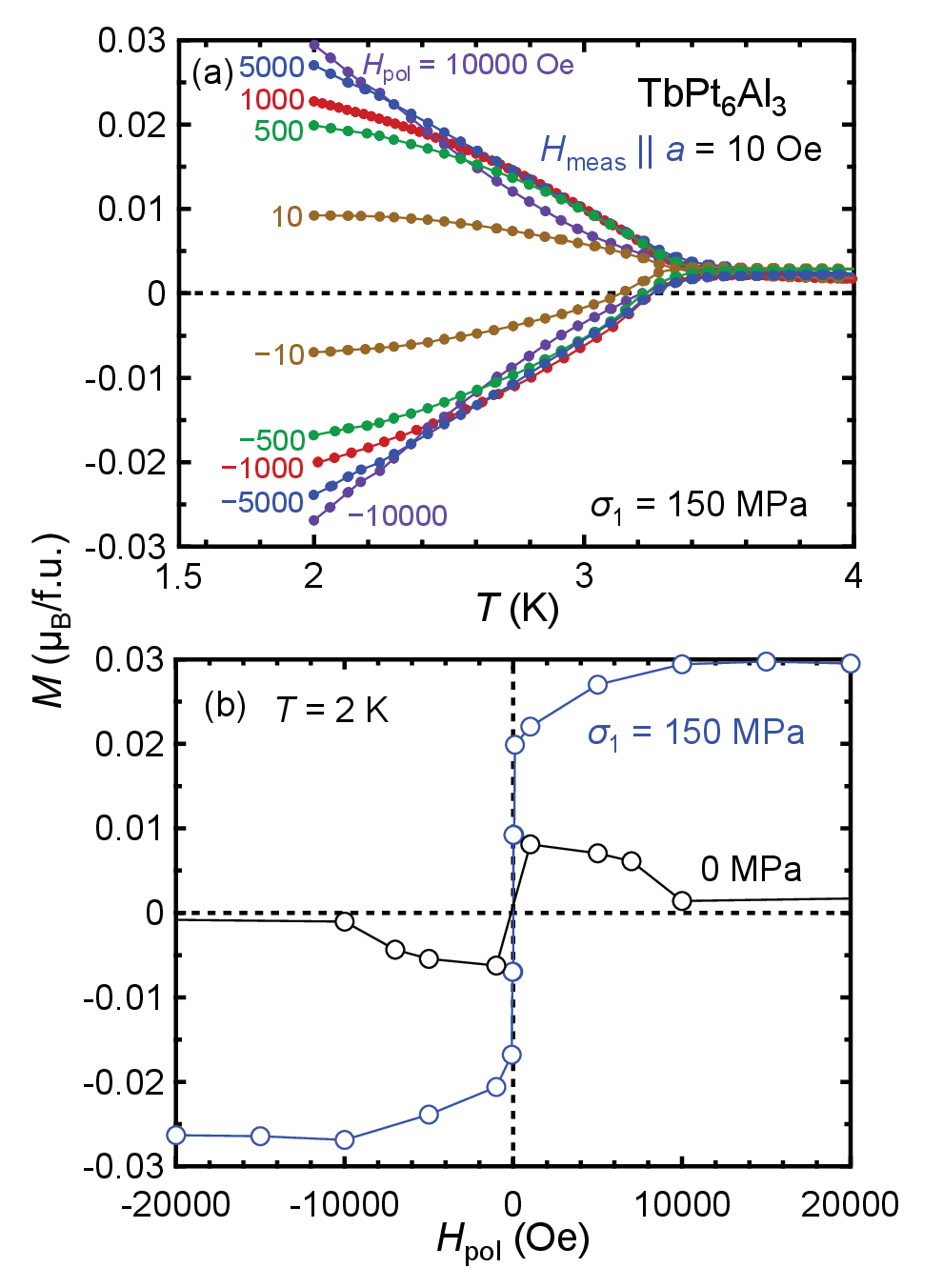}
	\caption{(Color online) (a) Temperature dependence of magnetization of $\rm TbPt_{6}Al_{3}$ at $\sigma_{1}$ = 150 MPa and $H_{\rm meas} \parallel a$ = 10 Oe under various poling fields $H_{\rm pol}$. (b) $H_{\rm pol}$ dependence of the value of $M$(2 K) at $\sigma_{1}$ = 0 and 150 MPa.}
	\label{f4}
	\end{center}
\end{figure}

The PZM coefficient is expressed by multipoles\cite{PG_multipole}.
Magnetic octupole $M_{3b}$, magnetic toroidal monopole $T_{0}$, and magnetic toroidal quadrupole $T_{u}$ are active in the magnetic point group of $\bar{3}m.1$ for $\rm TbPt_{6}Al_{3}$.
The PZM tensor components of $Q_{11}$ and $Q_{14}$ are represented by $M_{3b}$ and 2$T_{u}$, respectively.
Since the ZrO$_{2}$ piston in the present pressure cell moves vertically, it is difficult to investigate $Q_{14}$ by applied the shear stress of $\sigma_{4}$.

To realize the single-domain state, $M(T)$ measurements at $\sigma_{1}$ = 150 MPa and $H _{\rm meas}$ = 10 Oe were carried out at various poling fields $H_{\rm pol}$.
As shown in Fig. \ref{f4}(a), $M$ at 2 K decreases from 0.03 to 0.01 $\mu_{\rm B}$/f.u. with decreasing $H_{\rm pol}$ from 10000 to 10 Oe.
When $H_{\rm pol}$ is changed to a negative value of $-$10 Oe, $M$(2 K) becomes negative, corresponding to the reorientation of AFM domains.
Further decreasing $H_{\rm pol}$ to $-$10000 Oe, $M$(2 K) reaches $-$0.028 $\mu_{\rm B}$/f.u.
The $H_{\rm pol}$ dependence of the induced magnetization is understood by the population of the AFM domain\cite{PZM_MnTe}.
Figure \ref{f4}(b) represents the $H_{\rm pol}$ dependence of $M$(2 K) at $\sigma_{1}$ = 0 and 150 MPa.
The $M$(2 K)$-$$H_{\rm pol}$ curve at $\sigma_{1}$ = 150 MPa saturates to 0.03 $\mu_{\rm B}$/f.u. above $H_{\rm pol}$ of 10000 Oe, indicating that the single-domain state of $\rm TbPt_{6}Al_{3}$ is achieved at 10000 Oe.
The spontaneous moment observed at $\sigma_{1}$ = 0 MPa, on the other hand, is suppressed by increasing $H_{\rm pol}$.
For MnTe, $H_{\rm pol}$ = 1000 Oe was sufficient to control the AFM domain\cite{PZM_MnTe}.
Since the Tb$^{3+}$ ion has a total angular momentum $J$ = 6, the energy barrier for $\rm TbPt_{6}Al_{3}$ is larger than that for MnTe with a quenched orbital momentum of the Mn$^{2+}$ ion.

In conclusion, we have investigated the PZM effect in rare-earth-based $g$-wave AM $\rm TbPt_{6}Al_{3}$.
By measuring magnetization of single crystals under the uniaxial stress applied along the trigonal $a$ axis, we observed the linear development of the magnetic component below $T_{\rm N}$.
The PZM coefficient $Q_{11}$ is evaluated to be $9.1 \times 10^{-3} \rm \mu_{B}/(f.u.\,MPa)$ at 2 K, whose magnitude is larger than that for other AMs and AFMs with broken TRS.
This large PZM coefficient of $\rm TbPt_{6}Al_{3}$ confirms that strong relativistic SOC plays a crucial role in enhancement of the linear PZM effect.
Moreover, we have obtained the agreement of the critical component of $Q_{11}$ and the refined magnetic moment.
In order to gain insight into the mechanism of the large PZM response in rare-earth-based AMs and investigate $Q_{14}$ corresponding to $T_{u}$, measurements of the magnetostriction for $\rm TbPt_{6}Al_{3}$ and the PZM effect for the isostructural compound $R\rm Pt_{6}Al_{3}$ ($R$ = Nd, Sm, and Gd) are highly anticipated.

\begin{acknowledgment}
This work was supported by JSPS KAKENHI Grants No. JP22J20278, JP22KJ2336, JP22K03529, JP23H04870, JP25K00960, and JP26K17074.
This study was partly supported by Hokkaido University, Global Facility Center (GFC), Advanced Physical Property Open Unit (APPOU), funded by MEXT under "Support Program for Implementation of New Equipment Sharing System" (JPMXS0420100318).
The magnetization measurement and electron-probe microanalysis were performed at the Integrated Experimental Support/Research Division, N-BARD, Hiroshima University.
This work was performed under the Cooperative Research Program of "Network Joint Research Center for Materials and Devices (MEXT)".
We would like to thank T. Takabatake for his help in the crystal growth and discussions.
We acknowledge Y. Ogawa, S. Hayami, S. Ohara, F. Hori, and H. Harima for fruitful discussions about the piezomagnetic effect in the altermagnets.
\end{acknowledgment}

\end{document}